\documentclass[proceedings, preprint]{rmaa}
\usepackage{psfrag,color}

\SetYear{2010}
\SetConfTitle{LARIM}

\title{Radio emission from massive protostellar objects} 

\author{
  P. Benaglia\altaffilmark{1,2}}

\altaffiltext{1}{Instituto Argentino de Radioastronom\'{\i}a, 
CCT La Plata--CONICET, Villa Elisa, Argentina.}

\altaffiltext{2}{Facultad de Ciencias Astron\'omicas y 
Geof\'{\i}sicas, UNLP, Paseo del Bosque S/N, 1900, La Plata, Argentina 
(pbenaglia@fcaglp.unlp.edu.ar).}

\shortauthor{Benaglia}
\shorttitle{Radio emission from massive protostars} 
\listofauthors{P. Benaglia}
\indexauthor{Benaglia, P.}

\abstract{The study of the formation of massive stars present complex 
challenges from both theoretical and observational points of view. 
The initial phases of evolution, for instance, remain almost hidden except
at radio and IR wavelengths. In this article, 
after stating some of the problems of massive star formation, 
the role of radio observations to disclose the involved physics is
discussed. 
Historical observational findings are briefly outlined, and the connection
between low energy and high energy aspects of the phenomenon is 
addressed. Finally, 
data availability in the form of some new surveys is reported.}

\resumen{El estudio de la formaci\'on de estrellas masivas presenta
  a\'un hoy
complejos desaf\'{\i}os, tanto te\'oricos como observacionales. Las
fases m\'as tempranas son, en particular, de las que menos 
informaci\'on se tiene
ya que solo se detectan en longitudes de ondas milim\'etricas (mm) 
y sub-mm.  Aqu\'{\i} se describe qu\'e aportan los 
datos en radiondas, se resumen trabajos, y se 
discute la estrecha conexi\'on entre los procesos a bajas y altas energ\'{\i}as. 
Para terminar, se detallan brevemente fuentes alternativas de datos 
disponibles, a la luz de nuevos relevamientos.}

\addkeyword{H~II regions}
\addkeyword{ISM: Jets and outflows}
\addkeyword{Stars: Pre-main sequence}
\addkeyword{Stars: Mass loss}

\begin{document}
\maketitle


\section{Introduction}
\label{sec:intro}

Massive stars ($M_* \geq 8 {\rm M}_\odot$) are formed at the cores of 
dense molecular clouds with very low temperatures and very high  
densities (for a thorough discussion of the star-formation 
theories  the reader is referred to the recent review of McKee \&
Ostriker 2007).
These clouds remain undetectable except at radio and
near-IR wavelengths. It was not until the last decade that the angular 
resolution of the observing instruments improved significantly enough 
to allow a deep study of these almost-obscure regions. 
Figure~1 depicts a stellar-rich southern-sky star-forming region (SFR). 

\begin{figure}[!t]
  \includegraphics[width=\columnwidth]{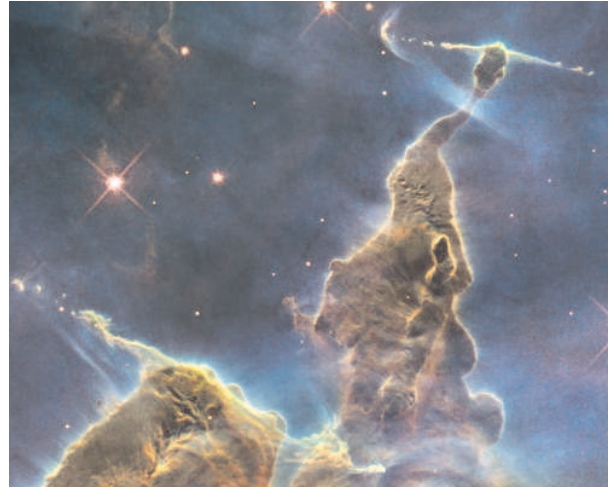}
  \caption{Star-forming pillars and protostellar objects at the Carina 
nebula. Credit: NASA, ESA, and M. Livio and the 
Hubble 20th Anniversary Team (STScI).}
  \label{fig:carina}
\end{figure}

The exact steps that lead to the formation of a high-mass star are 
not completely understood. The 
basic ingredients of the global process are the following ones. 
The gravitational collapse inside the cold cloud gives rise to an initial 
condensation that will become a protostar. Because of its angular momentum, 
the surrounding gas is accreted onto the protostar,
forming a disk. The twisting of the magnetic field that
pervades the region causes the protostellar object to eject matter, 
in the form of collimated polar jets of plasma (e.g. Arce et al. 2007). 
These jets propagate to large
distances (pc) and interact with the interstellar medium (ISM),
sweeping neutral gas in their path. 
Eventually, the piled up material stops the 
gas flow, forming a Herbig-Haro object. 
Only in the case of massive stars, the density of the central object 
is large 
enough to ignite nuclear reactions before the accretion comes to an end.

Figure 2 is a sketch of a star-forming molecular cloud, with
a variety of components. An extremely rich 
phenomenology can be appreciated.

\begin{figure*}[t]
\begin{center}
  \includegraphics[width=12cm]{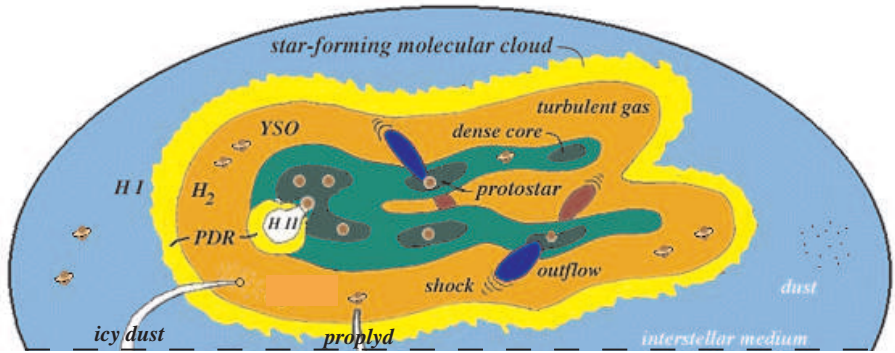}
  \caption{Sketch of a star-forming molecular cloud and components. YSO: young 
stellar objects, PDR: photo-dissociation regions, HI: neutral hydrogen gas,
${\rm H}_2$: molecular hydrogen gas, H II: ionized hydrogen regions, 
proplyds: disks with protoplanets (adapted from P. Myers, personal 
communication).}
  \label{fig:cloud}
\end{center}
\end{figure*}


Regardless of the protostellar mass value, there is generally a dense
core, jets and an accretion disk. 
Jets and disks were first discovered associated with low mass
protostars,
before Chini et al. (2004) detected for the 
first time an infrared disk around a massive protostar. 

\subsection{Complications at high-mass star formation}

High-mass star formation can be summarized in four phases: compression 
of cold cores, collapse of hot cores, accretion onto a massive 
protostellar object and disruption to give birth to main sequence 
stars. A description of what can be observed in each phase is 
given by Zinnecker and Yorke (2007).

Massive stars produce intense UV flux, orders of magnitude higher than 
low-mass stars. This flux photoevaporates and photodissociates the 
matter found in its way, creating ionized hydrogen regions. The radiation 
pressure, together with its effects, are also larger in massive
stars. There is disk 
dissipation, specially in the inner regions, close to the protostar. 
High-mass stars often form in non-single systems, where 
competitive accretion and n-body interactions are usually at work.

The issues just listed illustrate why high-mass star formation cannot be 
considered simply as a scaled-up version of low-mass star formation.
From the observational point of view, the scarcity of massive stars, 
their large distances, multiplicity 
and time scales of the changes they undergo 
preclude to get a complete picture
even with the best suited instruments.


\section{Motivation for radio studies}
\label{sec:motivation}

Basically, radio continuum data provide information on 
the source geometry and radiation regime. Line observations, besides 
probing the abundant neutral hydrogen, reveal gas distribution and 
kinematics in general, temperature, mass, density, opacity, etc.

Studies of maser lines are particularly important. Observations of 
several species account for estimates on different physical variables. For 
instance, ${\rm NH}_3$ and CS are emitted by dense molecular gas, generally 
surrounding the central source, and it might participate in driving gas
flows. ${\rm HCO}^+$ reports on the electron density in high-velocity gas. 
OH detections unveil the presence of low density gas in the outflow, it can be 
associated with high-mass star formation, where FIR radiation from 
heated dust is the pumping agent.
Water masers are indicative of young stars of low and high masses, 
and pinpoint molecular gas with $T \sim$  500K as in shocks 
(e.g. at dense clumps in winds). ${\rm CH_3OH}$ is related to
high-mass star formation and hot 
molecular cores besides traces dense gas in or near compact HII regions. 
${\rm CH_3CN}$ is also a disk tracer and presents low abundance in dense 
regions. SiO is an important print of molecular outflows; etc.

From the very beginning of the star-formation development, some physical 
processes leave signatures only at radio wavelengths, like the emission 
from cold dense cores. Radio waves can get through regions of dust 
almost without being absorbed. The spectral energy distribution built
from millimeter to sub-mm data helps to determine the stage of a 
massive young stellar object. Figure~3 shows an example of how much
information is provided 
by IR to radio observations towards a molecular cloud where stars are 
being formed (Pillai et al. 2006).

\begin{figure}[!t]
  \includegraphics[width=\columnwidth]{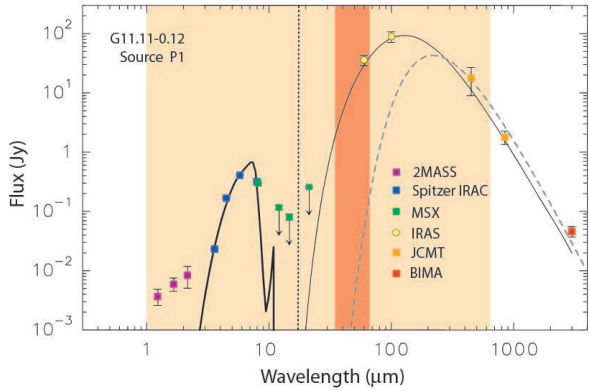}
  \caption{Low-frequency SED of the infrared dark cloud 
G11.11-0.12 (Pillai et
    al. 2006). The color of the symbols stands for the telescope used.
The sub-mm and mm fluxes appear to be produced by a 
protostar of 8 ${\rm M}_\odot$, 1200 ${\rm L}_\odot$ and $T = $15000K,
    with an envelope. The Spitzer-IRAC fluxes may represent a second
object. The NIR emission can be produced by scattering from tenuous gas 
above a disk.}
  \label{fig:g11.11}
\end{figure}

Magnetic fields play a fundamental role in massive-stars 
formation, not only in gravitational collapse but also during gas 
ejection. Measurements, however, remain a nontrivial task. Non-thermal 
spectral indices and polarisation measurements at radio frequencies constitute 
fundamental informants.

Radio data can be obtained all-day long, and continuous instrumental 
developments have allowed to attain improved sensitivity and 
similar angular resolution than at 
other energy ranges (e.g. VLBA, EUV, APEX, Spitzer, etc).

Of the high-mass star forming stages, cold and hot dark clouds 
are radio observables and, later, hyper compact, ultra compact and
regular ionized regions (Zinneker \& Yorke 2007, Beuther et al. 2007).


\section{Some discoveries and examples}
\label{sec:milestones}

In the context of star-forming regions and protostellar objects, the 
term radio jet commonly refers to a collimated flow of ionized gas, 
with velocities of $\sim$100 km s$^{-1}$, formed by the primary wind 
from the inner part of a star-disk system. On the other hand, gas of 
the environment, accelerated by interaction with the wind, forms 
lobe-shaped structures. This gas, moving at velocities of tens of km 
s$^{-1}$ represents the well-known molecular outflows (Bachiller 2007).

Radio observations of stellar jets and outflows were only possible
with the advent of modern radio interferometers. Some historical 
significant papers are reviewed below.

In 1980, both Snell et al. (1980) and Rodr\'{\i}guez et al. (1980) 
presented evidence of molecular outflows in star forming regions. The 
former group, 
of CO outflow lobes of $\sim$0.5 pc from L1551, a cloud that includes 
the objects HH~28, 29, and 102. The latter, through detection of 
high-velocity CO wings toward the SRF of Cep A, at 50km s$^{-1}$, 0.3 pc long. 
Pravdo et al. (1985) could image, for the first time, the 
ionized emission from a HH object (HH~1--2). 

Rodr\'{\i}guez et al. (1989) discovered the first non-thermal jet of a SFR
in Serpens. Later, Mart\'{\i} et al. (1993) reported on VLA 
observations of very collimated jets 
towards the luminous HH~80-81 ($\sim 2 \times 
10^4 {\rm L}_\odot$, $d = $1.7 kpc). 
Radio counterparts of HH~80 and 81 are 
aligned with an exciting source, about 2.3 pc to the south, linked by a 
patchy jet-like structure. A radio source named HH~80 North was 
discovered, symmetric to the corresponding southern one with respect to 
the central object. 
The spectral indices up to 10 GHz resulted positive 
($S_\nu \propto \nu^\alpha$) 
for the central source, and negative for HH~80, 
81, and 80 North. G\'omez et al. (2003), using
VLA data, determined the spectrum of the central source, and could detach the 
contributions of a protostellar disk and a collimated jet. Very 
recently, Qiu et al. (2009) were able to detect episodic molecular 
bullets ($\sim$100 km s$^{-1}$) that would be ejected close to the 
central source.

Observations of water masers were soon used to physically describe 
young stellar objects. Torrelles et al. (1996) mapped ${\rm H}_2{\rm 
O}$ emission, probably coming from a disk around a protostar with 
jets, in Cep A -- HW2 region. Very recently, new VLBI maser
observations (Torrelles et al. 2011), were performed over Cep A HW2. 
The data were taken at five epochs to measure proper motions of the 
maser spots. They found masers remaining static at 1 arcsec scale, but at 
shorter scales they draw an ellipse, supposedly around an unseen YSO.  The 
final results can be explained with two molecular outflows: one slow 
($\sim$50 km 
${\rm s^{-1}}$), at a wide angle ($100^0$), and a second collimated
($20^0$) fast gas 
jet ($\sim$500 km ${\rm s^{-1}}$). They also saw hints 
of a rotating wind around a central 
mass protostar of 20 ${\rm M}_\odot$.

Alcolea et al. (1993) have reported dramatic evidence for bipolar
flows too, based on the proper motions of water maser spots from a source
named ``TW'', close to W3(OH), from VLBI observations.
The results encouraged a study by Reid et al. (1995)
who used archive VLA data from 4 to 15 GHz to derive a 
non-thermal spectral index $\alpha$ for TW of $-0.6$, 
and size decreasing with frequency.
Wilner et al. (1999), by means of dedicated VLA observations at
8.4 GHz, confirmed the previous results, consistent with emission
produced by a synchrotron jet.

By the same time, Ray et al. (1997) could measure with MERLIN, at less
than $0.1''$ 
scale, circularly polarized emission from the outflows in the T Tau multiple
system. They derived magnetic field values of the order of several Gauss. 

By looking at L1551 IRS5, Rodr\'{\i}guez et al. (2003) discovered
at 3.5 cm binary jets, with slightly different orientations. They 
proposed a geometry of a binary system of protostellar objects. 
The complex source could be 
resolved by Lim \& Takakuwa (2006) using 7mm -- VLA + Pie Town 
data. They found the disks are 20 AU apart, and could 
modeled the disks and jets contributions.

In 2004 -- 5, disks around massive protostars became neatly 
detected in radio waves, like that of G24.78$+$0.08 (Beltran et al. 
2004). A detailed molecular study towards W51 North 
by Zapata et al. (2010) considered VLA and SMA 
data at various molecular lines (Fig. 4). The 
thesis that each molecule probes gas at a different temperature (and 
consequently radius) seems to be confirmed by observations. This 
allowed to propose a scenario like the one portrayed in Figure 4. 

\begin{figure}[!t]
  \includegraphics[width=\columnwidth]{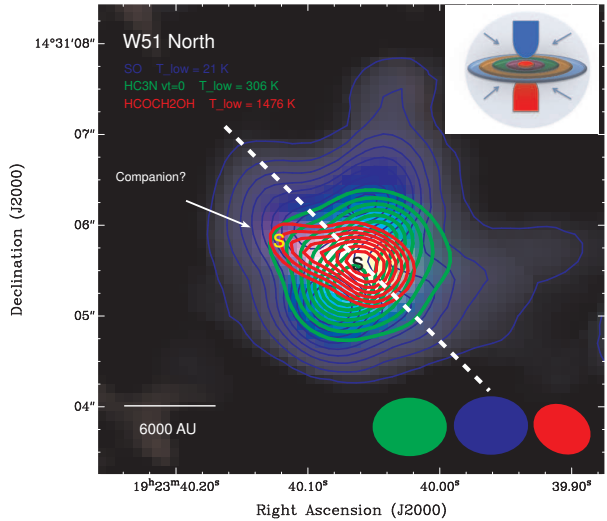}               
  \caption{Molecular emission from a circumstellar large disk 
around the massive protostar W51 North.
SO (blue), HC$_3$N (green) and HCOCH$_2$OH (red) emission. 
Synthesized beams are on the right bottom corner. 
Upper-right inset: diagram showing the molecular layers
of the disk, still surrounded by an infalling envelope.  
A clear temperature gradient is
observed across the disk.}
  \label{fig:mol-layes}
\end{figure}


Regarding specifically high-mass protostellar objects,
Garay et al. (2003) discovered the first very luminous ($6.2 \times 
10^4 {\rm L}_\odot$) young stellar object with non-thermal jets: the 
source IRAS 16357$-$4247. At 2.9 kpc, it was observed from 1 to 25 GHz 
(Brooks et al. 2007 and references therein) radio continuum and CO. 
It has a core mass of 1000 ${\rm M}_\odot$, and a CO outflow of 
100 ${\rm M}_\odot$. The proper motion of the jet knots 
was explained assuming precession (Rodr\'{\i}guez et al. 2008).

IRAS 16353--4635 is an example of a star forming region for which radio 
plus near IR data can be used to separate protostellar sources 
and classify them according to mass (Benaglia et al. 2010). 
Figure 5 shows a spectral index map, built using 17 and 19 GHz
ATCA data,
superposed with near-IR observations. Complementary NIR spectroscopy 
allowed to identify a low 
and a high-mass protostars. Analysis of radio continuum emission
from 1.4 to 20 GHz suggested a possible outflow at the radio peak.

\begin{figure}[!ht]
  \includegraphics[width=6.7cm,angle=-90]{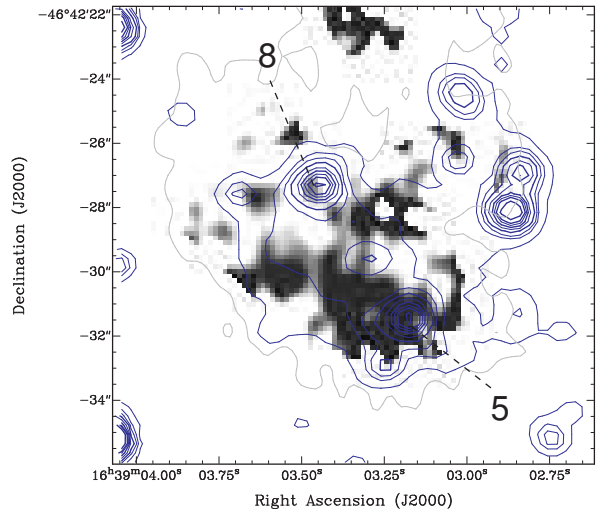}
  \caption{Spectral index map of IRAS 16353--4635, derived for
17.3~GHz and 19.6~GHz ATCA data, in grayscale: black is $-$1, white is 
$+$1. NTT $K_{\rm s}$-band emission is superimposed as blue 
contours from 25 to 400 Jy.
Grey contour: 3$\sigma$ level emission at 17.3 GHz. 
NTT sources 5 (low-mass protostar) and 8 (high-mass protostar) are 
marked (see Benaglia et al. 2010 for details).}
  \label{pbenaglia-4}
\end{figure}

A major breakthrough in the last years was the detection 
of strong linearly-polarized emission from a massive protostellar jet. 
Carrasco-Gonz\'alez et al. (2010) could measure from VLA-5 GHz data,
polarized light up to a degree of 30\%, coming from the jet of Herbig-Haro 
object \#80.

\begin{figure}[!t]
  \includegraphics[width=\columnwidth]{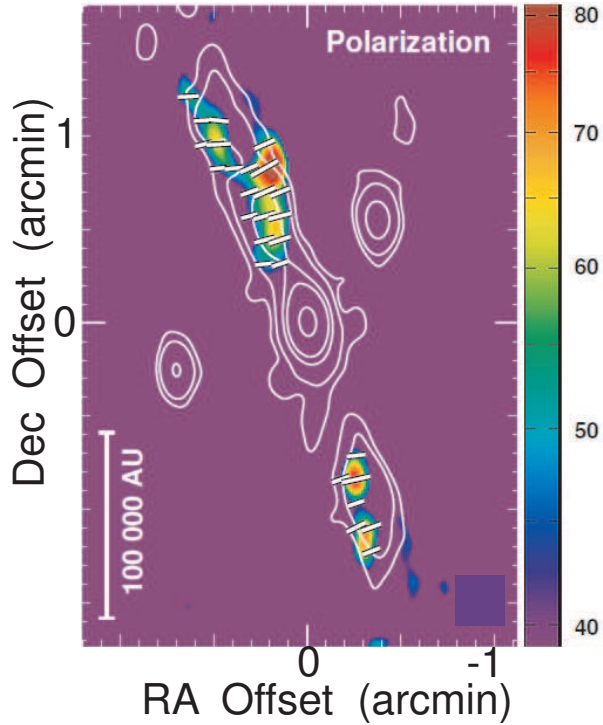}
  \caption{
Image of the HH~80-81 jet region at 6~cm wavelength. Linearly polarized 
continuum intensity image (color scale, units of $\mu$Jy~beam$^{-1}$). 
Polarization direction is shown as white bars. The total continuum 
intensity is also shown (contours; levels from 40 to 
3300 $ \mu$Jy~beam$^{-1}$).}
  \label{pbenaglia-f5}
\end{figure}


\section{The radio-gamma connection}

The high degree of linearly polarized light coming from HH~80 confirms 
the synchrotron nature of the radiation, produced by relativistic 
electrons.
The clear detection of relativistic particles in young stellar objects 
exposes the possibility of high-energy radiation in this kind of systems.
There are a bunch of mechanisms by which, both relativistic electrons 
and protons accelerated at the jets or in their terminal regions, can 
give rise to high-energy emission.

Araudo et al. (2007) discuss inverse Compton 
interactions between the same relativistic electrons involved in the 
radio emission, and the infrared photon field of the molecular cloud 
that hosts the massive young stellar object.

The electrons can also cool through relativistic Bremsstrahlung. In both 
cases, the maximum energies are of the order of 1 TeV. If the acceleration 
process for environmental protons is effective, then inelastic 
collisions with the nuclei of the cloud can yield pions. Neutral 
pions decay producing gamma rays. Charged pions inject secondary pairs, 
which in turn can contribute to the synchrotron radiation observed in 
radio waves. Since the losses are less severe for protons, the radiation 
produced in processes at which they are involved can reach 
energies $\sim$ 10 TeV.

The various processes that can lead to gamma-ray emission inside a 
star-forming region are discussed in detail in Romero (2008, 2010).

A complete model of the source HH~80 has been recently developed by 
Bosch Ramon et al. (2010). It can explain the measured non-thermal radio 
emission but also predict a possible further detection with large 
Cherenkov arrays like CTA. Figure 7 shows the spectral energy
distribution as it is expected from HH~80 (Bosch Ramon et al. 2010).

These works open a new window for the study of the star-formation 
processes, based on the detection of radiation produced by 
high-energy phenomena. In this way, gamma-ray astronomy can be used to 
probe the physical conditions in star-forming regions and particle 
acceleration processes in the complex environment of massive molecular 
clouds.

\begin{figure}[!t]
  \includegraphics[width=\columnwidth]{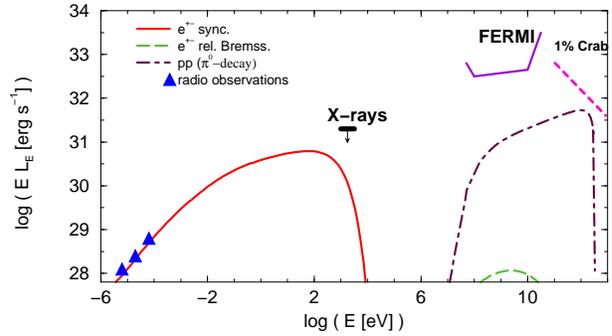}             
  \caption{
Spectral energy distribution of the non-thermal emission for HH~ 80.  
Observational points are from Mart{\'\i} et al. (1993). The X-ray 
detection is shown as an upper-limit. The {\it Fermi} line stands for 
1~yr/5~$\sigma$ sensitivity. The curve above 100~GeV corresponds to 0.01 
~Crab (Bosch Ramon et al. 2010).}
  \label{fig:radiogamma}                 
\end{figure}

 
\section{Surveys}
\label{sec:surveys}

The description of powerful forthcoming instruments like the SKA, 
ALMA, LOFAR and the numerous ways in which they will contribute  
 in broadening the comprehension of the 
processes already mentioned have been widely covered. Instead, 
I would like to close this review outlying additional
 sources of information like radio surveys, that are focused on the earlier
 evolutionary phases of high-mass stars. A 
few projects were chosen, by way of examples.

\paragraph {\bf The Galactic Census of High and Medium-mass Protostars} 
or CHaMP\footnote{http://www.astro.ufl.edu/~peterb/research/champ/} 
(Barnes et al. 2005). The survey consists of performing systematic 
observations with the Nanten Telescope at the lines of CO, $^{13}{\rm
  CO}$, 
${\rm C^{18}O}$  
and ${\rm HCO^+}$, to identify massive dense molecular clumps within giant 
molecular clouds, and follow-up observations with the Mopra telescope at 
higher resolution. To begin with, the area covered is delimited by 
$300^0 > l > 280^0$, $-4^0 < b < +2^0$. Complementary 
observations at the $J$, $H$ and $K_{\rm s}$ 
bands with the AAT, and other molecular 
line observations will also be implemented. The goal is to learn about the physical 
processes that dominate the star formation activity. The large number of 
clouds surveyed will provide information on a population of protostars, 
and derive lifetimes and other physical parameters.

\paragraph {\bf The Red MSX Sources Survey} or 
RMS\footnote{http://www.ast.leeds.ac.uk/RMS/}. This program (Urquhart
et al. 2011 and references therein) 
systematically searches all sky for massive young stellar objects, 
using far-IR data -where the spectral index distribution 
peaks-, and measurements of maser lines. A first selection of 2000 
candidates is the target of an extensive multi-wavelength campaign to 
confirm the YSO nature or eliminate confusing objects.
A major result so far is a web-based database of candidates, 
with 6-cm high-resolution continuum images, and Spitzer, MSX, 2MASS maps,
maser line data, etc.

\paragraph {\bf The MeerGAL Survey}\footnote{www.ska.ac.za/meerkat/index.php}, 
by M. Thompson and cols. 
It is planned to be the deepest and highest resolution 14 GHz survey
of the southern Galactic Plane. The observations will be conducted
using the African SKA pathfinder MeerKAT. 
Some specific tasks to achieve are: to carry out a
survey of hypercompact HII regions, to measure the thermal flux of
massive stars, and to perform methanol maser observations. The angular
resolution will be less than 1 arcsec, and the 
sensitivity below 0.1 mJy. It is expected to be finished by 2016.\\

{\small

\noindent Acknowledgements. It is a pleasure to thank the 
organizers of the meeting LARIM 2010:
the LOC which efficiently took care of every detail, 
and the SOC for the kind invitation. Special thanks to Luis Felipe
and Yolanda, to Enrique Vazquez-Semadeni and to Adriana Gazol.
P.B. is supported by ANPCyT (PICT 2007--00848), CONICET (PIP
2009--0078), and FCAG--UNLP (Proyecto G093).
}


\end{document}